\documentclass[aps,prx,groupedaddress,reprint,showpacs,pdftex]{revtex4-1}
\usepackage{graphicx}
\usepackage{tabularx}
\usepackage{dcolumn}
\usepackage{bm}
\usepackage{amsmath}
\usepackage{amssymb}
\usepackage{txfonts}
\usepackage{url}
\usepackage[usenames,dvipsnames]{color}
\definecolor{Gray}{gray}{0.0}
\definecolor{lightGray}{gray}{0.35}
\begin{document}
\title{
    Feature space of XRD patterns constructed by auto-encorder
}

\author{
  Keishu Utimula$^1$,
  Masao Yano$^2$, 
  Hiroyuki Kimoto$^2$, 
  Kenta Hongo$^{3,4,5}$, 
  Kousuke Nakano$^{6}$,
  Ryo Maezono$^{6}$ \\}

\affiliation{\\
  $^1$School of Materials Science, JAIST, 
  Asahidai 1-1, Nomi, Ishikawa 923-1292, Japan\\
  \\
  $^2$Toyota Motor Corporation, 
  1, Toyota-cho, Toyota, Aichi 471-8572, Japan\\
  \\
  $^3$Research Center for Advanced Computing 
      Infrastructure, JAIST, Asahidai 1-1, Nomi, 
      Ishikawa 923-1292, Japan\\
  \\        
  $^4$Center for Materials Research by Information Integration,
  Research and Services Division of Materials Data
  and Integrated System,
  National Institute for Materials Science,
  Tsukuba 305-0047, Japan\\
  \\
  $^5$PRESTO, JST, Kawaguchi, Saitama 332-0012, Japan\\
  \\
  $^6$School of Information Science, JAIST, 
  Asahidai 1-1, Nomi, Ishikawa 923-1292, Japan
}

\vspace{10mm}
\affiliation{$^{*}$
  mwkumk1702@icloud.com
}
  
\date{\today}
\begin{abstract}
It would be a natural expectation that 
only major peaks, not all of them, 
would make an important contribution 
to the characterization of the XRD pattern. 
We developed a scheme that can identify 
which peaks are relavant to what extent 
by using auto-encoder technique 
to construct a feature space for the XRD 
peak patterns. 
Individual XRD patterns are projected 
onto a single point in the two-dimensional 
feature space constructed using the method.
If the point is significantly shifted 
when a peak of interest is masked, 
then we can say the peak is relevant 
for the characterization represented 
by the point on the space. 
In this way, we can formulate the relevancy 
quantitatively. 
By using this scheme, we actually found 
such a peak with a significant peak intensity 
but low relevancy in the characterization 
of the structure. 
The peak is not easily explained 
by the physical viewpoint such as 
the higher-order peaks from the same plane index, 
being a heuristic finding by the power of 
machine-learning. 
\end{abstract}
\maketitle

\section{Introduction}
\label{sec.intro}
{\it Materials informatics} (MI) becomes 
to form a major topic as an 
application of {\it big data science} 
for the materials discovery~\cite{Ikebata2017,2018OSE}. 
In the topic, some efforts are found 
to utilize auto-encoder technique, 
{\it e.g.},
applying it to reduce noise
in the Laser Induced Breakdown Spectroscopy (LIBS) spectra. 
~\cite{2018SHI}
The auto encoder technique~\cite{2006HIN}
was originally introduced 
as a method for compressing the dimensions 
of an image data using neural networks.~\cite{2006HIN} 
The technique has since been applied to 
extractions features of data~\cite{2008RUS}, 
noise eliminations for image data~\cite{2013CHO}, 
detections of signal anomalies~\cite{2014SAK}, 
and to data generations to form 
images {\it etc.}~\cite{2018HUA}

\vspace{2mm}
One of the promising topics in the MI is 
the recognitions of the peak patterns 
of X-ray diffraction~(XRD) analysis,
~\cite{2014KUE,2017SUR,Iwasaki2017,2018SHA,2018STA,2018XIN,2019UTIb}, 
which is common in crystal structure analysis 
in experiments~\cite{Hongo2018} 
In terms of how to capture the features of patterns, 
all the previous studies above are based on the application 
of time-series signal recognition techniques. 
The Auto-encoder technique, on the other hand, 
takes a different approach to capture features: 
Putting features is regarded as a grouping 
operation made over all data, each of which has 
some data dimension ({\it e.g.,} 
the number of data on time-axis). 
The grouping can be seen as a coarse-graining 
of information, and hence corresponds to 
the compression of data dimension by some means. 
In the auto-encoder concept, 
such a compression is realized by the 
neural network technique~\cite{2006HIN}. 
There has been several recent reports 
applying the neural network to 
the problems in materials science
~\cite{2018DIM, 2018NAS}.

\vspace{2mm}
In the present study, 
we applied the auto-encoder technique 
using a neural network 
to the recognition of the XRD spectrum 
to extract features of 
peak patterns. 
The features are well captured 
on a two-dimensional space (feature space) 
from which we can reproduce the original 
data again precisely. 
The distance on the space can 
recognize the compositions of 
the doped compounds. 
Surprisingly the distance 
is also capable to be used 
to distinguish which peaks are relevant for 
capturing the features. 
This is quite in contrast to 
the traditional manner for 
researchers in materials science, 
interpreting the 
patterns from so called 
'crystallographic viewpoints', 
where the features are 
captured from {\it multiple} natures 
such as the Bravais lattice estimations 
({\it e.g.}, from Rietveld analysis~\cite{1969RIE}), 
indexing of each diffraction peak, 
fitting of lattice constants 
({\it e.g.}, Vegard's law~\cite{1921VEG, 1991DEN}), 
and charge density estimations 
({\it e.g.}, from MEM analysis ~\cite{1993JAU, 1985WEI, 1987GUL, 1990SAK}). 
Though our achievement would just 
be described as 'a dimensional 
compression from original 11,900-dimensional 
into two-dimensional by an auto-encoder' 
in a technical reporting manner, 
it would be a matter of surprising 
that the target problem which is 
traditionally considered from 
{\it multiple view} of human 
is now well captured only in two-dimensional 
by machines. 

\section{Model and Methodology}
\label{sec.det.spec}
We applied our framework to 
a data set of XRD patterns 
of magnetic alloys, 
[Sm$_{(1-y)}$Zr$_y$]~Fe$_{12-x}$Ti$_x$, 
with different concentrations $(x,y)$.~\cite{2019UTIb} 
A fixed concentration still includes 
multiple possibilities on the 
inequivalent locations of substituent sites.~\cite{2019UTIb}
For 10 different concentrations considered here (See Table ~\ref{table.searching}), 
we have 150 XRD patterns in total.
All the XRD data are generated by 
simulations using density functional theory (DFT) 
which perform geometrical optimization 
being in quite well coincidence with 
experimental data.~\cite{2019UTIb}

\begin{table}[h]
  \caption{
    The numbers of inequivalent configurations of
    Sm$_{(1-y)}$Zr$_y$Fe$_{12-x}$Ti$_x$ to be considered.  
    The numbers in brackets indicate the structures constructed
    from the $2\times2\times2$ supercell (Sm/Zr),
    while the rest are constructed from the $2\times2\times1$ 
    supercell (Fe/Ti).}
\begin{center}
  \begin{tabular}{c|rrrrr}
    $y\backslash x$ & 0.0 & 0.5 & 1.0 & 1.5 & 2.0 \\
    \hline
    0.000 & 1 & 13 & 22 (1) & 27 & 61 \\
    0.125 & -  & -  & (2) & - & - \\
    0.250 & -  & -   & (7) &- & - \\
    0.375 & -  & -   & (6) &- & - \\
    0.500 & -  & -   & (10) &- &-  \\
  \end{tabular}
\end{center}
\label{table.searching}
\end{table}

\vspace{2mm}
The data are give to 
a neural network as 
11,900-dimensional vector components, 
$\left\{ I(2\theta_j)\right\} _{j=1}^{11,900}$
(2$\theta=0\sim$120~deg., 
$\delta\theta$ = 0.01~deg.) 
at the input layer.
Note that the range of $2\theta$ in our case 
corresponds to the experimental setting 
using synchrotron radiation 
($\lambda = 0.496 \AA$).
As an auto-encoder, the network 
{\it encodes} an input vector,  
compressing its dimension by 
hidden layers, finally to two-dimensional, 
and then decodes it toward the output-layer 
with the same dimension as that of input.  
The parameters in the network 
are optimized so that the output could 
reproduce the input identically. 
For the implementation, 
we used PyTorch~\cite{2019PAS}
with activation functions, 
ReLU~\cite{2011GLO}
for hidden layers both for the encoder 
and the decoder. 
At the final layer of encoder (decoder), 
tanh (linear function)
is used as the activation function. 
Parameters are optimized by Adam~\cite{2014KIN}
algorithm combined with the error estimations 
by MSELoss(Mean Squared Error function)
and the L2-norm weight decay.
90\% of the data are used for the training 
and the rest are for the test. 
We found the optimal construction of 
the network (hyper-parameters) 
to minimize the mean error 
over ten samples for the numerical stabilization, 
being (the number of hidden layers) = 3, 
(Minibatch size) = 10, 
and (the number of epoch) = 1,000. 

\begin{figure}[htbp]
  \includegraphics[width=\hsize]{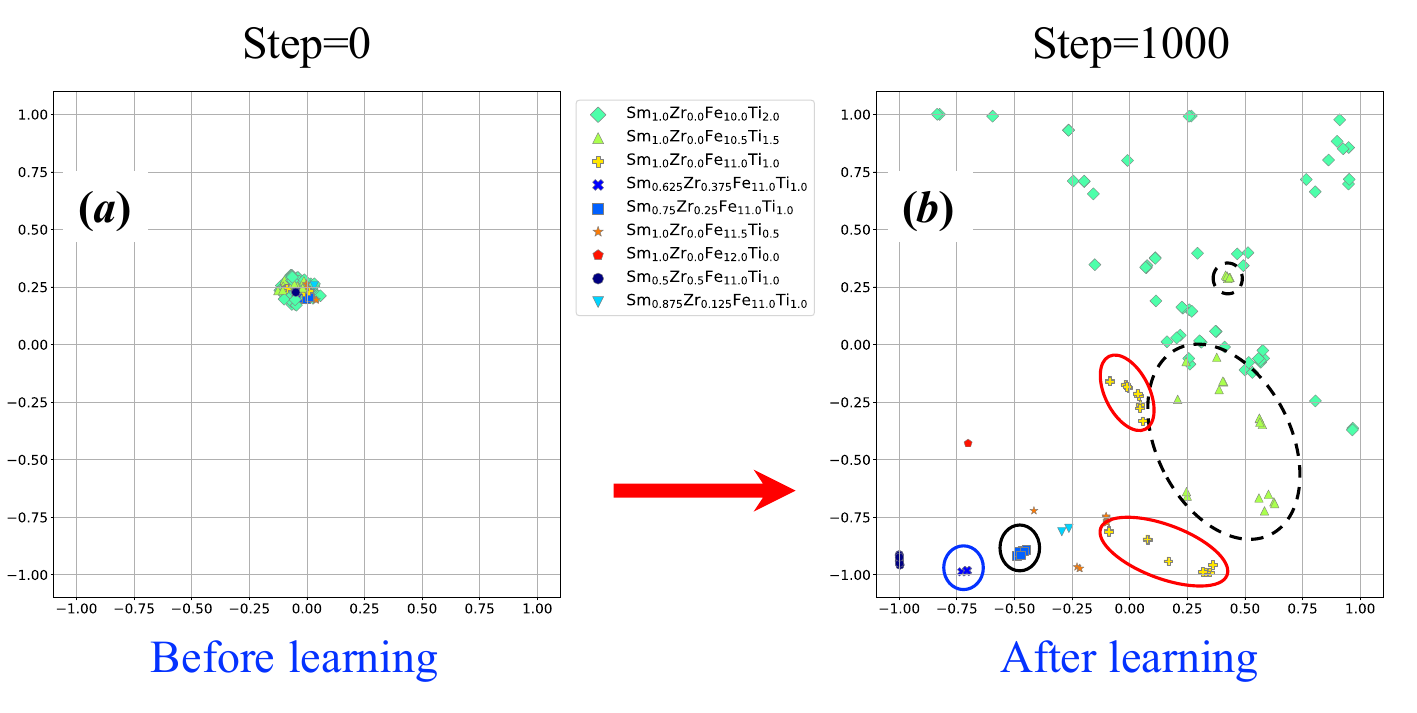}
  \caption{
Distributions of 150 XRD patterns projected to 
two-dimensional feature space composed by 
our auto-encoder. 
The points with the same symbol 
share the same concentration of 
atomic substitutions 
but differs in the location of 
the substituents being inequivalent 
in group theoretical manner.~\cite{2019UTIb} 
As the parameters are optimized 
({\it i.e.}, the learning of the neural 
network is completed), 
the initial distribution [panel (a)] 
gets to form clusters as shown in 
the panel (b), as enclosed by circles.
      }
  \label{afterLearning}
\end{figure}

\vspace{2mm}
The network squeezes the dimension of 
data 11,900-dimensional at the input 
into 128 $\to$ 64 $\to$ 12-dimensional
via three hidden layers and finally 
composes two-dimensional feature space 
at the final edge of encoder 
(namely, at the middle of 
the whole auto-encoder). 
Fig.~\ref{afterLearning} shows 
the feature space, 
on which 150 XRD patterns are 
projected to each point. 
The initial distribution of the points 
[panel (a)] gets to be scattered 
to form clusters as shown in the 
panel (b) as the learning in the network 
proceeds. 
We observe that 
each cluster [as shown 
by circles in the panel (b)]
is formed by the patterns sharing 
the same concentration [corresponding to 
the same symbol] of substituents, 
and hence that 
the feature space could 
well recognize the compositions of 
samples. 

\section{Results and discussion}
\label{sec.results}
\begin{figure}[htbp]
  \includegraphics[width=\hsize]{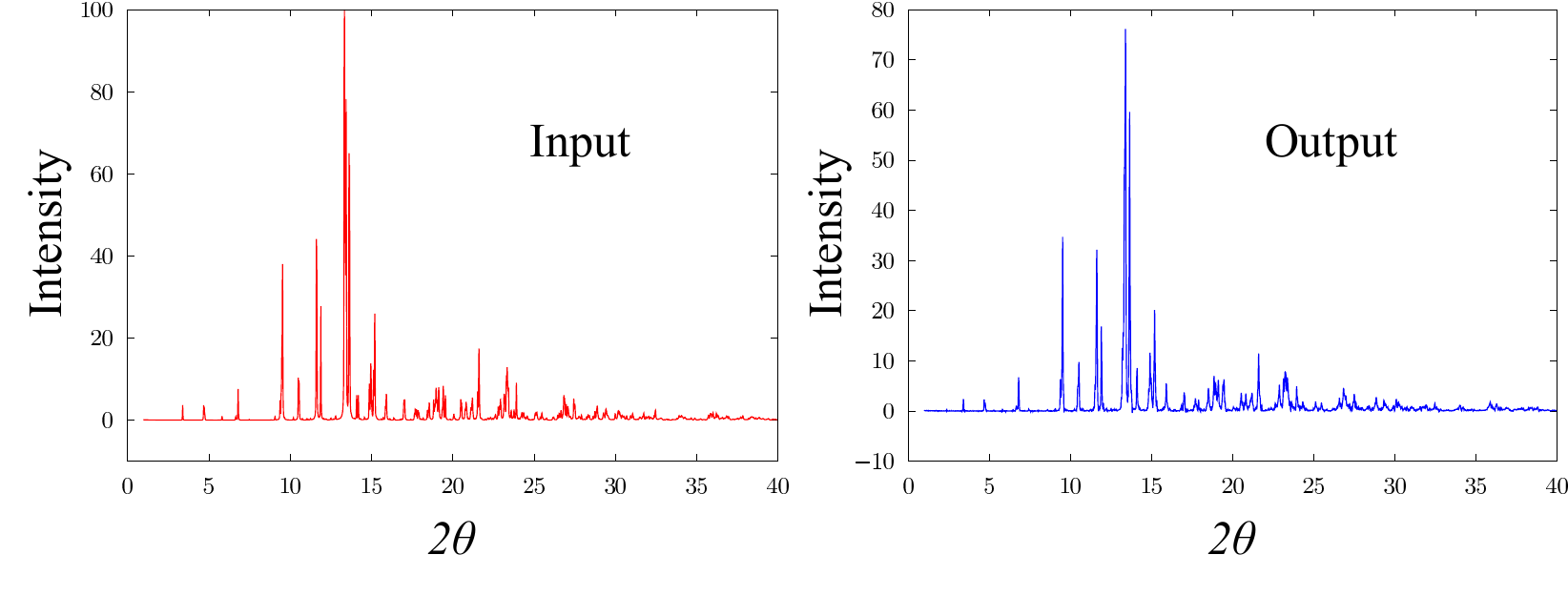}
  \caption{
The comparison between the input 
XRD pattern and 
the reproduced one by our 
auto-encoder. 
They coincide well each other 
at the level of human eyes resolution, 
and hence ensuring 
data learning on the auto-encoder 
being well achieved.
Note that the range of $2\theta$ in our case 
corresponds to the experimental setting 
using synchrotron radiation 
($\lambda = 0.496 \AA$).
  }
  \label{output3}
\end{figure}

Fig.~\ref{output3} shows 
the comparison between the input 
XRD pattern and 
the reproduced one by the 
auto-encoder. 
They coincide well each other 
at the level of human eyes resolution, 
namely that of the traditional analysis 
by materials science researchers. 
The result is hence ensuring that 
the data learning of XRD patterns 
on the auto-encoder is well 
achieved.

\vspace{2mm}
In the following discussions, 
we provide several possible ideas 
how to utilize the extracted 
features of XRD patterns, 
(A) identifications of doping concentrations 
for a given XRD pattern of unknown samples, 
(B) clarifying the {\it irrelevancy} of 
each peak in a pattern in contributing 
to the features. 
(C) generating artificial XRD patterns 
for a given concentration
as the {\it interpolation} over XRD patterns
to omit expensive {\it ab initio} analysis.

\subsection{Identification of concentrations}
Since in Fig.~\ref{afterLearning}(b) 
the closer samples in the feature 
space have the closer concentration, 
it is quite likely for the samples 
with unknown concentrations 
to be projected to the location 
being closer to the sample having 
the closer concentration. 
One could then estimate the unknown 
concentration from the distance 
on the feature space. 

\begin{figure}[htbp]
  \includegraphics[width=\hsize]{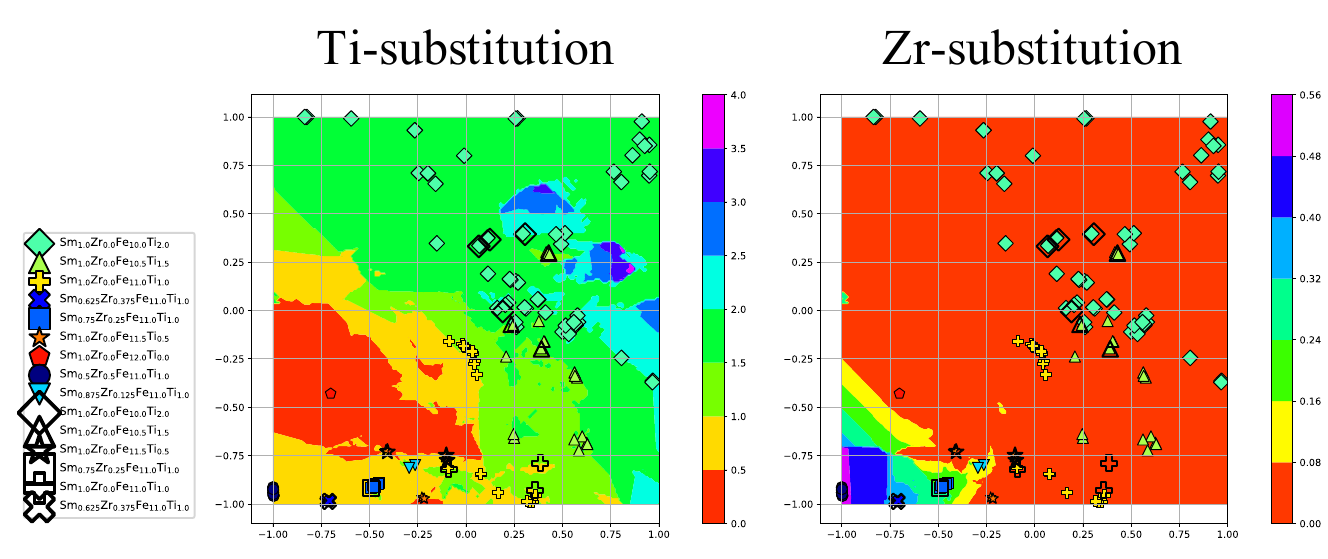}
  \caption{
Estimations of the substitutional 
concentrations for any points 
on the feature space 
represented by color maps. 
By using such maps, we can 
identify the concentration for 
a given sample to be projected 
on a point in the space. 
  }
  \label{map}
\end{figure}

\vspace{2mm}
For such an identification, 
we {\it paint} the feature space like Fig.~\ref{map}
so that the color could correspond 
to the concentration. 
From the color to where a given sample 
is projected, we would be able to estimate 
the concentration of the sample. 
One would wonder the painting 
can be performed by 
the clustering technique (unsupervised 
machine-learning), such as k-means method. 
\cite{1967MAC}
However, it comes to the conclusion 
that at least k-means method is never 
the appropriate choice for the painting: 
the method relies on the concept of 
'center of gravity' of the data 
forming each cluster region. 
The data is sorted based on the distance 
from the center of gravity of each cluster. 
Such concept would work well 
if each cluster forms a simply-connected region, 
otherwise the closer distant points 
from the center might be out of the region. 
As seen in the solid circled region in 
Fig.~\ref{afterLearning}(b), our data 
could form such clusters with the same 
feature being not simply-connected 
(triangle symbols form two regions 
separated each other). 
Rather than using global 
knowledge over the data such as 
the center of gravity, 
it is better to use local 
information in the vicinity of 
the projected point in the space, 
as concluded. 

\begin{figure}[htbp]
  \includegraphics[width=\hsize]{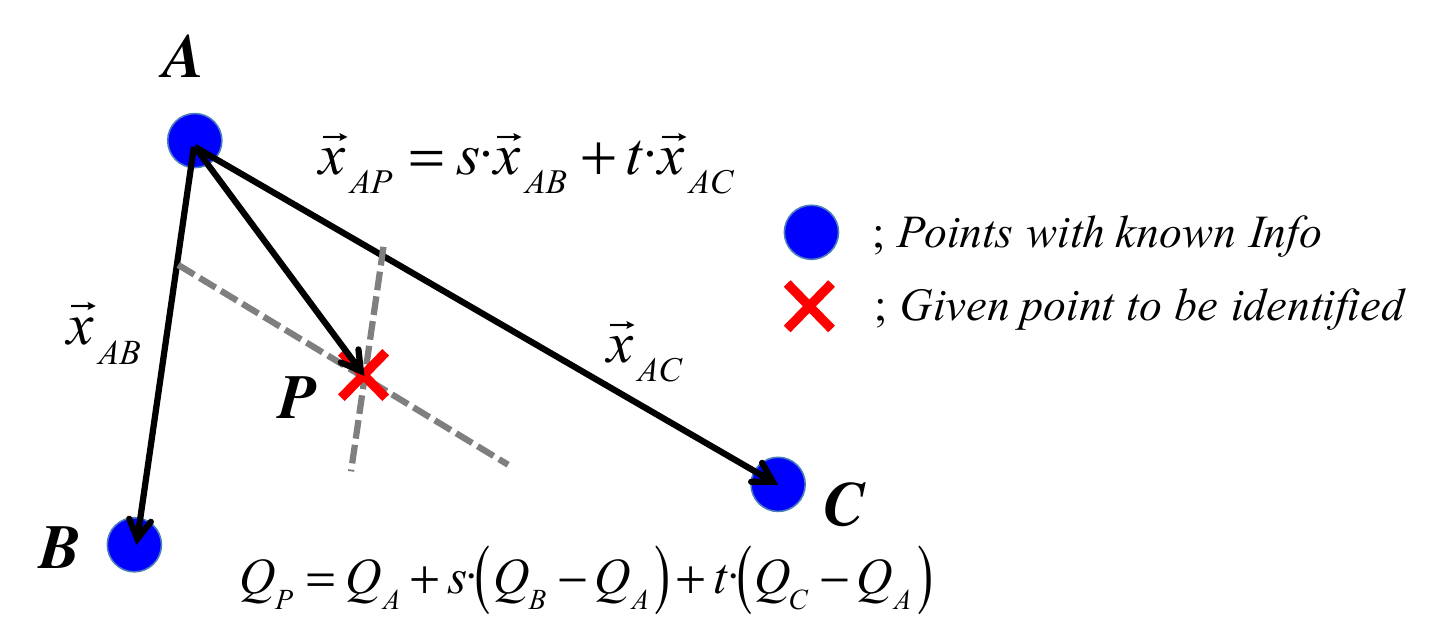}
  \caption{
    Inference of unknown quantity for a given 
      point $P$ in the feature space. 
      It simply estimates the quantity $Q_P$ 
      by using the fraction of $s$ and $t$ 
      measured by the distance in the space 
      assuming the linear interpolation. 
      }
  \label{interpolation}
\end{figure}

\vspace{2mm}
As a simplest implementation to use 
such a local information, 
we could use the linear interpolations, 
as explained in Fig.~\ref{interpolation}. 
Suppose the target sample 
(an XRD pattern) 
with an unknown property (concentration in 
the present case) is projected to $P$, 
in the vicinity of which 
we could three data, $A$, $B$ and $C$, 
with known properties, $Q_A$ {\it etc.}. 
When the location of $P$ is 
described as 
\begin{eqnarray}
{{\vec x}_{AP}} = s\cdot {{\vec x}_{AB}} 
    + t\cdot {{\vec x}_{AC}} \ , 
\end{eqnarray}
the quantity for $P$ is naively 
estimated using the same fractions as
\begin{eqnarray}
{Q_P} = {Q_A} + s\cdot\left( {{Q_B} - {Q_A}} \right) 
              + t\cdot\left( {{Q_C} - {Q_A}} \right) \ . 
\end{eqnarray}
For the way how to assign three 
known points in the vicinity of $P$ into 
$A$, $B$ and $C$, respectively, 
we would choose them so that 
the inner products $(\vec x_{AB}\cdot x_{AP})$
and $(\vec x_{AC}\cdot x_{AP})$ may 
get larger to obtain plausible 
interpolations (as is shown 
in Fig.~\ref{interpolation}). 
The condition may interpreted that 
$P$ should be located inside 
of the triangle formed by 
$A\sim C$, leading to 
the condition for the fractions, 
$s,t>0$ and $1>s+t>0$, 
as implemented in a program. 

\vspace{2mm}
By sweeping $P$ over everywhere 
in the feature space with picking up 
three nearest points, 
we can estimate the concentration 
of any point on the space by using 
the above formalism, getting 
the {\it painted} map as shown in 
Fig.~\ref{map}. 
By using this map, we can 
identify the concentration for 
a given sample to be projected 
on a point in the space. 
Using test data points (for which 
the concentration is known), 
we can examine the performance 
of the prediction by the map 
by comparing the known answer 
to the estimation. 
Table.~\ref{estimation} shows 
the list of the results 
in order of the poorest grades in 
the error which is defined 
for a concentration 
Sm$_{c_1}$Zr$_{c_2}$Fe$_{c_3}$Ti$_{c_4}$ 
between the answer (A) and 
the estimation (E) as 
\begin{equation}
\delta = \sum\limits_{\alpha  = 1}^4 
{{{\left( {c_\alpha ^{\left( E \right)} 
- c_\alpha ^{\left( A \right)}} \right)}^2}} \ . 
\label{CompDist}
\end{equation}

\begin{table}[htb]
  \caption{[estimation]
    The list of the estimation performance 
    of sample concentrations for test data 
    using our linear estimation, 
    shown in order of the poorest grades in 
    the error which is defined 
    in Eq.~(\ref{CompDist}) [The worst ten is shown]. 
    The worst score is coming from 
    the fact that there is little 
    leaning data for this case as 
    explained in the main text.
  }
  \begin{tabular}{l|l|l}
   \hline
   \hline
    True composition & Estimated composition & Error\\
    \hline
    Sm$_{1.000}$Zr$_{0.000}$Fe$_{11.000}$Ti$_{1.000}$ & Sm$_{1.000}$Zr$_{0.000}$Fe$_{10.733}$Ti$_{1.267}$  & $1.43 \times 10^{-1}$\\
    Sm$_{1.000}$Zr$_{0.000}$Fe$_{11.000}$Ti$_{1.000}$ & Sm$_{1.000}$Zr$_{0.000}$Fe$_{10.958}$Ti$_{1.042}$  & $3.59 \times 10^{-3}$\\
    Sm$_{1.000}$Zr$_{0.000}$Fe$_{10.500}$Ti$_{1.500}$ & Sm$_{1.000}$Zr$_{0.000}$Fe$_{10.522}$Ti$_{1.478}$  & $9.85 \times 10^{-4}$\\
    Sm$_{1.000}$Zr$_{0.000}$Fe$_{11.000}$Ti$_{1.000}$ & Sm$_{1.000}$Zr$_{0.000}$Fe$_{11.022}$Ti$_{0.978}$  & $9.31 \times 10^{-4}$\\
    Sm$_{1.000}$Zr$_{0.000}$Fe$_{10.500}$Ti$_{1.500}$ & Sm$_{1.000}$Zr$_{0.000}$Fe$_{10.490}$Ti$_{1.510}$  & $1.85 \times 10^{-4}$\\
    Sm$_{1.000}$Zr$_{0.000}$Fe$_{11.500}$Ti$_{0.500}$ & Sm$_{0.999}$Zr$_{0.001}$Fe$_{11.494}$Ti$_{0.506}$  & $7.39 \times 10^{-5}$\\
    Sm$_{1.000}$Zr$_{0.000}$Fe$_{11.500}$Ti$_{0.500}$ & Sm$_{1.001}$Zr$_{-0.001}$Fe$_{11.505}$Ti$_{0.495}$ & $4.80 \times 10^{-5}$\\
    Sm$_{0.625}$Zr$_{0.375}$Fe$_{11.000}$Ti$_{1.000}$ & Sm$_{0.623}$Zr$_{0.377}$Fe$_{11.000}$Ti$_{1.000}$  & $1.03 \times 10^{-5}$\\
    Sm$_{1.000}$Zr$_{0.000}$Fe$_{10.500}$Ti$_{1.500}$ & Sm$_{1.000}$Zr$_{0.000}$Fe$_{10.501}$Ti$_{1.499}$  & $4.26 \times 10^{-6}$\\
    Sm$_{0.750}$Zr$_{0.250}$Fe$_{11.000}$Ti$_{1.000}$ & Sm$_{0.749}$Zr$_{0.251}$Fe$_{11.000}$Ti$_{1.000}$  & $1.20 \times 10^{-6}$\\
\hline
\hline
  \end{tabular}
  \label{estimation}
\end{table}

\vspace{2mm}
Looking at the errors shown 
in Table.~\ref{estimation}, 
the achievement of the prediction 
is fairly well, getting around less than 
0.5\% in general except one with 
the worst accuracy ($\delta_{\rm worst}^{(1)}$ = 0.14). 
We could find out that the case 
is attributed to the shortage of the 
available data, which is the common 
problem when the machine-learning is applied 
to materials science. 
In the present case, 
the available number is completely 
specified 
by the number of elements 
for an equivalent location of substituents 
in the space group theoretical manner, 
~\cite{2019UTI}
without further variations because 
the XRD patterns in our study are all generated 
by zero-temperature {\it ab initio} simulations. 
For more practical and realistic applications
to be compared with experimental data, 
one would generate further samples 
with more variety using, {\it e.g.} 
molecular dynamics at finite temperature 
starting from the optimized geometry at $T=0$ 
to put thermal fluctuations. 

\vspace{2mm}
The general consensus is that it is impossible 
to make sense of what is meant by a quantity 
of vertical/horizontal axis in such a 
compressed two-dimensional feature space. 
In the present case, however, 
it may be possible to trace 
the meaning to some extent 
by the following way: 
The features in the present XRD case are 
anyway those characterizing crystal structures 
tuned by the substituents, 
{\it e.g.}, the $c/a$ ratio {\it etc.}
It is straightforward to generate 
such variations to capture such features 
artificially by the simulations. 
By observing how the locations of the 
projected points on the feature space 
are affected by such artificial/typical 
change in the structure, 
we could extract the trend what 
the vertical/horizontal axis represents. 

\subsection{Identifying relevant peaks}
It would be a fundamental question 
related to XRD that "Do we really need 
all the peaks to characterize the structure 
with XRD patterns? Only a specific bandwidth 
of $2\theta$ matters?" 
In this context, we would like to identify 
how much each peak is relevant in  
characterizing the feature of XRD. 
The most naive idea to measure the 
relevancy would be to find out 
disappeared peaks after an XRD 
is reproduced by the auto-encoder 
by comparing between input and output 
patterns. 
Such an idea is not 
working unfortunately 
because, as seen in Fig.~\ref{output3}, 
the optimization of our neural network 
is so successful that the output 
can reproduce an input pattern very well 
(no such disappeared peaks found). 

\begin{figure}[htbp]
  \includegraphics[width=\hsize]{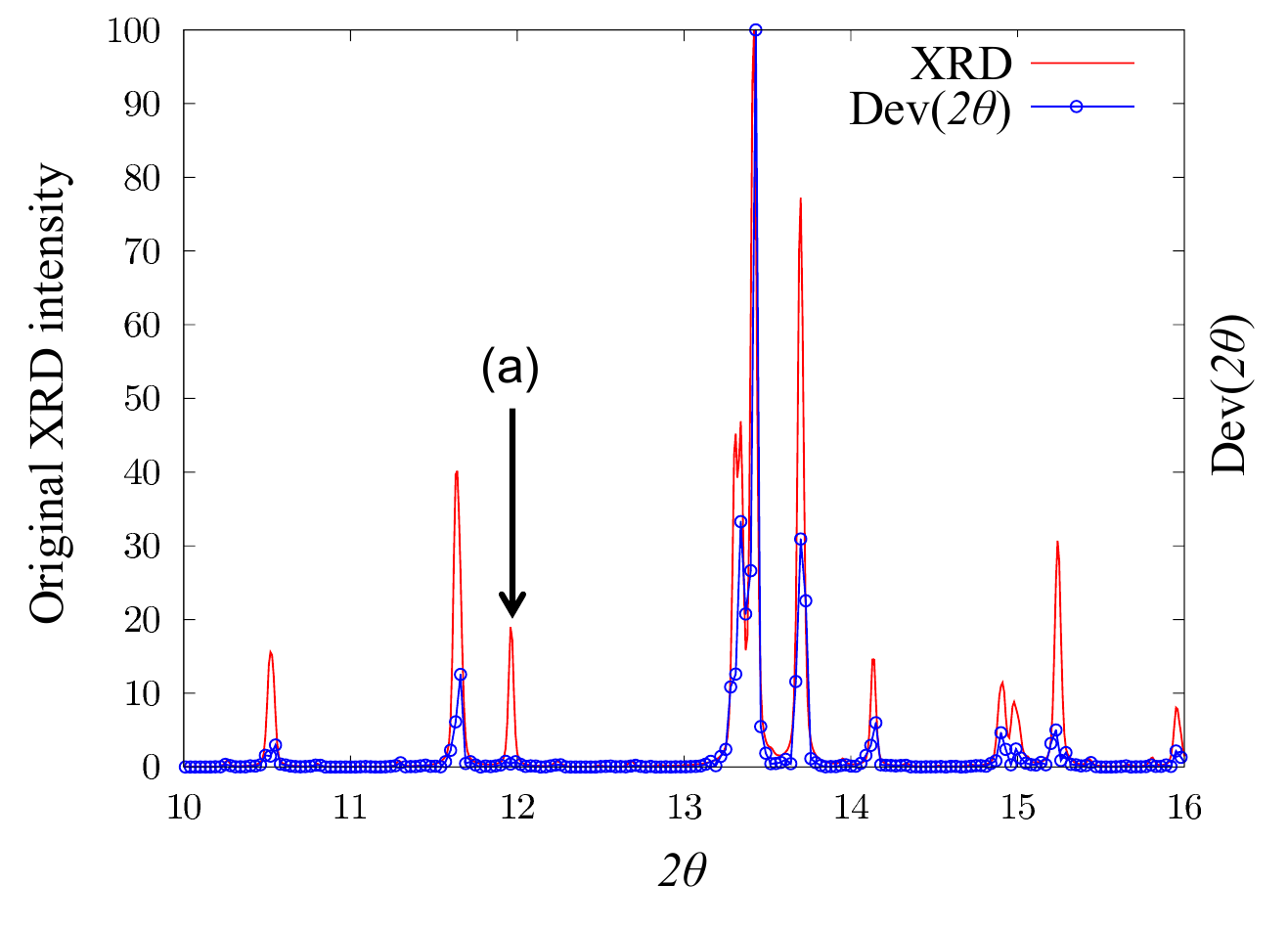}
  \caption{
  The intensity of the relevance 
  of each peak as defined by ${\rm Dev}(2\theta)$ 
  (open circles) compared with the 
  original XRD pattern (solid line).
    This XRD pattern is for test data,
    the composition is Sm$_{1.0}$Zr$_{0.0}$Fe$_{10.5}$Ti$_{1.5}$.
  As a natural expectation, they are well 
  correlating but at some peaks [{e.g.}, at (a)], 
  they are not. 
}
  \label{dev}
\end{figure}

\vspace{2mm}
We can instead take such an idea 
that the relevancy would be measured 
by how much the projected location 
on the feature space is affected 
when the peak considered is masked. 
Here we define a mask vector $M(2\theta)$
having the same dimension as XRD's
as its components being zero
for $2\theta\sim 2\theta +\Delta$ 
$\left(\Delta = 0.03\right)$ and
one for otherwise.
By using the mask vector, 
the masked XRD pattern can then be represented as 
${\rm XRD}_M(2\theta) := M(2\theta) \circ {\rm XRD}$,
where the symbol $\circ$ stands for a Hadamard product.
Suppose $\mathcal{P}\left[{\rm XRD}\right]$ to be
the projected XRD onto the feature space,
the displacement on the feature space caused by masking 
can be evaluated as a normal distance on the feature space:
${\rm Dev}(2\theta)=
\left|\mathcal{P}\left[{\rm XRD}_M(2\theta)\right]
-\mathcal{P}\left[{\rm XRD}_0\right]\right|$.
This would measure how much 
the masked peak at $2\theta$ affects 
the location on the feature space, 
and hence correspond to the 
'intensity of relevancy'.
The plot of the intensity, ${\rm Dev}(2\theta)$, 
is shown in Fig.~\ref{dev}, 
superimposed on the pattern of XRD.
This XRD pattern is for test data,
the composition is Sm$_{1.0}$Zr$_{0.0}$Fe$_{10.5}$Ti$_{1.5}$.

\vspace{2mm}
As a natural expectation, both 
the original XRD and the relevancy 
are well correlating 
(namely the relevancy takes the large value 
when the XRD intensity gets larger), 
but at some peaks [{e.g.}, at (a) 
in Fig.~\ref{dev}], they are not. 
Though the peak at (a) has significant 
intensity, its relevance is found to be 
almost zero. 
One might come to the idea why this 
might happen that such a peak might 
come from a higher-order plane index, 
and then has some intensity but 
the relevant information is already 
reflected to the lower-order ones. 
We found, however, that 
the peak (a) has a fundamental 
index, (0,0,1), being not a 
higher-order. 
The finding of (a) is hence 
a sort of "beyond the human's expectation", 
and can be said to be the knowledge 
that can only be obtained by machine-learning.

\subsection{interpolation of XRD patterns}
XRD provides inferences not only of the 
lattice constants 
but also of other valuable information 
such as strains, crystalline sizes 
{\it etc.} 
It is, therefore, desirable that 
learning data to be compared with 
given experimental XRD have 
finer resolution in the composition 
as possible. 
Realizing such a finer resolution 
by {\it ab initio} simulations 
is, however, generally very difficult. 
Conventional treatment of atomic substitutions 
using supercell model requires 
a very large supercell to represent 
tiny percentage of substitutions, 
which is practically impossible to be performed. 
That's how we came up with the idea of 
"interpolating" discrete points 
(50\%, 25\%, 12.5\%, {\it etc.}) 
that are feasible at realistic cost 
in {\it ab initio} simulations. 
Such an conceptual idea can actually be 
realized by our feature space. 
Since a contour for composition is constructed 
on the feature space as in Fig.~\ref{map}, 
we can interpolate to find 
a point corresponding to a desired 
composition on the contour. 
The output pattern generated by the auto-encoder 
projected from the point on the feature 
space could be a plausible XRD pattern 
for the composition, that is obtained 
without costly {\it ab initio} simulations. 
This approach can be applied not only to 
XRD but also to other spectrum and even 
to other physical quantities to be 
evaluated by {\it ab initio} simulations. 
The approach would also be regarded as another 
remedy to overcome the long-standing difficulties 
in {\it ab initio} calculations, 
{\it i.e.}, the problem of the computational 
cost for treating tiny concentrations 
with larger supercell models.

\section{Conclusion}
\label{sec.conc}
We developed an auto-encoder to form a feature space 
describing XRD patterns. 
The framework has well been trained by 
the data to reproduce input peak patterns 
at the level of human eye's recognition. 
Each XRD is projected to a point on 
the two-dimensional feature space, forming 
clusters with concentrations being 
closer each other. 
The distance in the space can be used 
to estimate the concentrations 
of any given samples of XRD by 
projecting it on the space. 
We could compose a contour map 
on the space describing the concentration 
by using a linear interpolation 
connecting between the training data. 
By examining the prediction performance 
using test data, we confirmed 
that it achieves around less 
than several percent except the cases 
with little training data. 
We proposed a couple of interesting 
applications of the feature space: 
The way to identify the {\it relevancy} for 
each peak on characterizing XRD features 
is proposed. 
The idea is implemented as the 
observation of location shifts 
of a point on the feature space 
when a concerned peak is masked 
from the original XRD pattern. 
By this method, we found 
a non-trivial case with a peak 
having a considerable intensity 
but little relevancy for which 
we could not make reasonable 
account for the tiny relevancy 
from the physics viewpoint 
({\it e.g.}, higher-order 
reflections {\it etc.})
As another application, 
we proposed how to {\it interpolate} 
XRD patterns to avoid expensive 
{\it ab initio} simulations with 
the difficulty to handle tiny changes 
in concentrations. 
The interpolation can be made 
on the feature space and hence 
the auto-encoder can generate 
an artificial but plausible XRD pattern 
for the interpolated point with 
desired composition. 
The approach would be regarded as 
a useful remedy to achieve 
finner resolutions of concentrations 
with avoiding the computational cost 
when handled by {\it ab initio} 
simulations. 

\section{Acknowledgments}
The computations in this work have been performed 
using the facilities of 
Research Center for Advanced Computing 
Infrastructure at JAIST. 
R.M. is grateful for financial supports from 
MEXT-KAKENHI (19H04692 and 16KK0097), 
from FLAGSHIP2020 (project nos. hp190169 and hp190167 at K-computer), 
from Toyota Motor Corporation, from I-O DATA Foundation, 
from the Air Force Office of Scientific Research 
(AFOSR-AOARD/FA2386-17-1-4049;FA2386-19-1-4015), 
and from JSPS Bilateral Joint Projects (with India DST). 
The X-ray diffraction (XRD) measurements
for the Sm-Fe-Ti system were performed
at beamline BL02B2 of SPring-8
under the proposal No. 2017B1634.

\bibliographystyle{apsrev4-1}
\bibliography{references}
\end{document}